\documentclass[aps,prl,twocolumn,superscriptaddress,showpacs]{revtex4}

\usepackage{graphicx}

\begin{document}

\title{Quantum Criticality in an Organic Magnet}

\author{M. B. Stone}
 \affiliation{Condensed Matter Sciences Division, Oak Ridge National Laboratory,
Oak Ridge, Tennennesse 37831, USA}
 \affiliation{Department of Physics and Astronomy, Johns Hopkins
University,
Baltimore, Maryland 21218}
\author{C. Broholm}
 \affiliation{Department of Physics and Astronomy, Johns Hopkins
University,
Baltimore, Maryland 21218}
 \affiliation{National Institute of Standards and Technology, Gaithersburg,
Maryland 20899}
\author{D. H. Reich}
 \affiliation{Department of Physics and Astronomy, Johns Hopkins
University,
Baltimore, Maryland 21218}
\author{O. Tchernyshyov}
 \affiliation{Department of Physics and Astronomy, Johns Hopkins
University, Baltimore, Maryland 21218}
\author{P. Vorderwisch}
\affiliation{Hahn-Meitner Institut, D-14109 Berlin, Germany}
\author{N. Harrison}
\affiliation{National High Magnetic Field Laboratory, LANL, Los Alamos, New
Mexico 87545}

\date{\today}

\begin{abstract}

Exchange interactions between $S=\frac{1}{2}$ sites in piperazinium
hexachlorodicuprate produce a frustrated bilayer magnet with a
singlet ground state. We have determined the field-temperature phase
diagram by high field magnetization and neutron scattering
experiments. There are two quantum critical points: $H_{c1}=7.5$ T
separates a quantum paramagnet phase from a three dimensional,
antiferromagnetically-ordered state while $H_{c2}=37$ T marks the
onset of a fully polarized state. The ordered phase, which we describe
as a magnon Bose-Einstein condensate (BEC), is embedded in a quantum
critical regime with short range correlations.  A low temperature
anomaly in the BEC phase boundary indicates that additional low
energy features of the material become important near $H_{c1}$.

\end{abstract}


\maketitle The concept of a critical transition between different
phases of matter at temperature $T=0$ is central to many complex
phenomena in strongly correlated systems \cite{sachdevbook}. Quantum
critical points (QCPs) give rise to anomalous properties through a
range of temperatures, and may be responsible for heavy fermions
\cite{abfermionreview}, non-fermi-liquids \cite{colemanrev}, and the
anomalous normal state of doped cuprates \cite{varma}. Among the
non-thermal tuning parameters accessible to the experimentalist,
doping has been applied to access QCPs in
heavy fermion intermetallics \cite{schroeder,steglich} and copper
oxide superconductors \cite{tranquada}, and hydrostatic pressure has
been used to expose anomalous superconducting \cite{saxena} and
metallic \cite{grosche} phases in weak itinerant magnets. While
magnetic fields generally induce conventional transitions between
states with static spin order, exceptions are found in anisotropic
spin systems where a transverse magnetic field, $H$, can drive a
transition from spin order at $H=0$ to a quantum disordered state
\cite{bitko}. The reverse transition from a quantum paramagnet (QP) in
zero field to an anisotropic ordered state in high fields has been
observed in certain organo-metallics \cite{Eckert79,cuhpcl,ndmap}.
While materials with such behavior are often quasi-one-dimensional,
recent experiments have revealed a wider range of cooperative
phenomena in higher dimensional
systems\cite{cavadini,stonecuhpcl,jaime2004}. Owing to the
simplicity of the low energy Hamiltonian, high field experiments on
organo-metallic magnets are a promising route to new information
about quantum criticality.

We provide a comprehensive analysis of the $H-T$ phase diagram of a
quasi-two-dimensional (2D) frustrated organo-metallic
antiferromagnet (AFM) with two field driven QCPs.
Key results include a detailed characterization of a Bose-Einstein
condensation (BEC) in the vicinity of a zero-temperature quantum critical
point.  We also find a low $T$
anomaly in the BEC phase boundary, which may indicate that nuclear spins
and/or phonons are important thermodynamic degrees of freedom close to
the QCP.

\begin{figure}
\centering\includegraphics[scale=0.66,angle=-90]{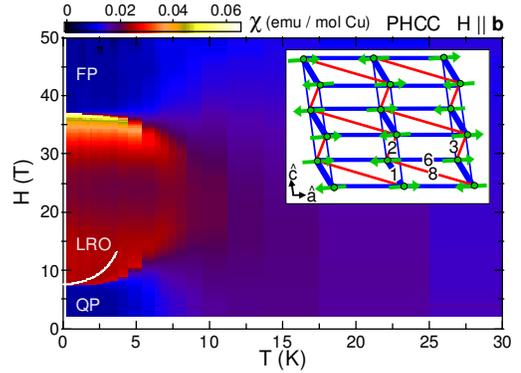}
\caption{\label{fig:phcccontour}(Color online) Differential susceptibility $\chi (H,T)$ for
PHCC. Solid white line for $H < 14.2 $ T is the the line of
phase transitions defined by the onset of N\'{e}el order at
higher field.  The terms QP, LRO and FP are explained
in the text.  Inset: PHCC structure
showing the Cu$^{2+}$ $S=\frac{1}{2}$ sites (solid circles) viewed
along the {\bf b} axis. Interacting spins are connected by lines
with thickness proportional to the contribution to the $H=0$ ground state
energy. Red[blue] bonds are frustrated[unfrustrated] and
increase[decrease] the ground state energy. Numbering corresponds to
Ref.~\protect\cite{stonephccprb01}. Vectors show the ordered spin
structure at $T=1.65$ K and $H=13.7$ T.}
\end{figure}

Experiments were carried out on the quasi-2D $S=\frac{1}{2}$ quantum
AFM piperazinium hexachlorodicuprate
($\rm(C_{4}H_{12}N_{2})Cu_{2}Cl_{6}\equiv $ PHCC). The crystal
structure is composed of Cu-Cl sheets in the {\bf a}-{\bf c} plane,
separated by piperazinium layers \cite{stonephccprb01,battaglia88}.
Magnetic properties are dominated by the Cu-Cu interactions within
individual sheets shown in Fig.~\ref{fig:phcccontour}. The magnetic
connectivity is that of an oblique bilayer, with the strongest bond, \textit{i.e.} the dimer, bond 1,
providing interlayer coupling. Frustrated interlayer bonds 2
and 8 may also play a role in producing a singlet ground state
with strong correlations to five near neighbors. Magnetic
excitations at $H = 0$ are dominated by a dispersive triplet of magnons, also known as the triplon, with a
bandwidth $W=1.8$~meV and an energy gap $\Delta = 1$~meV. Cluster
expansion analysis of the $H=0$ dispersion indicates that the
strongest intra-layer bond $J_6 \approx 0.36 J_1$, while the
frustrating bonds $J_2,J_8$ are $\approx 0.1 J_1$
\cite{martinmullerpreprint}. An experimental limit of $0.2(1)$ meV
has been placed on the out of plane dispersion and the triplons are
degenerate to within 0.05 meV.

Magnetic susceptibility measurements were performed at the National High Magnetic Field Laboratory
using a compensated-coil susceptometer in pulsed fields up to $H=50$
T for $0.46$ K $\le T \le 30$ K. The sample was a 1.36 mg
hydrogenous single crystal with $\mathbf{H} \parallel$~{\bf b}.  Elastic
neutron scattering measurements were performed on the FLEX
spectrometer at the Hahn-Meitner Institut (HMI).  The sample was composed of two 89\% deuterated
single crystals with total mass 1.75 grams, coaligned within
0.5$^{\circ}$ and oriented in the $(h0l)$ scattering plane,
$\mathbf{H}\parallel$~{\bf b}. A room temperature graphite filter or a liquid
nitrogen cooled beryllium filter was employed in the scattered beam
for neutron energies 14.7 and 2.5 meV respectively. Beam divergence
was defined by the $^{58}\rm Ni$ neutron guide before the
monochromator and 60$^{\prime}$ collimators elsewhere.

\begin{figure}
\centering\includegraphics[scale=0.46,trim = 0in 0.1in 0in
0in]{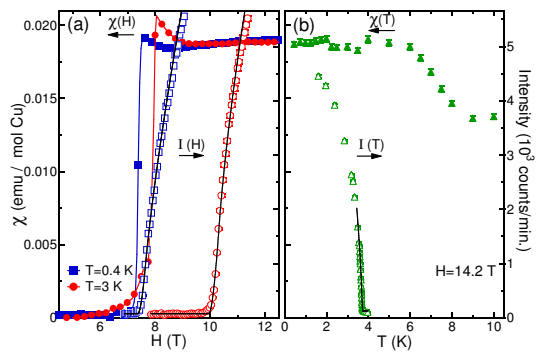}
\caption{\label{fig:phccbraggint} (Color online) (a)
Field-dependence of $\chi(H)$ (filled symbols) at $T = 0.46$ K and 3
K, and of the $(\frac{1}{2}0\bar{\frac{1}{2}})$ AFM Bragg intensity
$I(H)$ (open symbols) at $T = 0.42$ K and 2.85 K. The onset of 3D
ordering occurs above $H_{c1}$. Solid lines are fits to $I(H)$
described in text. (b)  Temperature-dependence of
$(\frac{1}{2}0\bar{\frac{1}{2}})$ Bragg peak at $H = 14.2$ T,
compared to $\chi(H,T)$.}
\end{figure}

Differential magnetic susceptibility data, $\chi(H,T)=dM/dH$, are
shown in Fig.~\ref{fig:phcccontour}. At  $T=0.46$ K there is
evidence for two quantum transitions from gapped phases with
$\chi=0$ for $H< H_{c1} \approx 7.5$ T and $H> H_{c2} \approx 37$ T
to a magnetizable state in the intermediate field range. $\chi(H)$
at the lower transition is shown in Fig.~\ref{fig:phccbraggint}(a).
Integrating $\chi(H,T=0.46\ {\rm K})$ yields a saturation
magnetization of $1.097(3)\mu_B$ per spin, identifying the high
field phase as fully spin-polarized (FP).

In the intermediate field phase AFM Bragg peaks were found at wave
vectors ${\bf Q} = {\bf \tau} +(0.5,0,0.5)$ where ${\bf \tau}$ is a
reciprocal lattice vector of the chemical cell. The lower bound on
the order parameter correlation length in the $\bf a-c$ plane is
$2.0(2)\times 10^3$ \AA. Analysis of peak intensities yields the
spin structure in Fig.~\ref{fig:phcccontour}, which is consistent
with bond energies measured in the zero field phase in that
(un)frustrated bonds correspond to (anti)parallel spins. Normalizing
to incoherent scattering and assuming long range order (LRO) along $\bf b$ yields
$g\mu_B\langle S\rangle =0.33(3)g \mu_B$ at $T = 1.65$ K and $H =
13.7$ T.

\begin{figure}
\centering\includegraphics[scale=0.61,trim = 0in 0.1in 0in
0in]{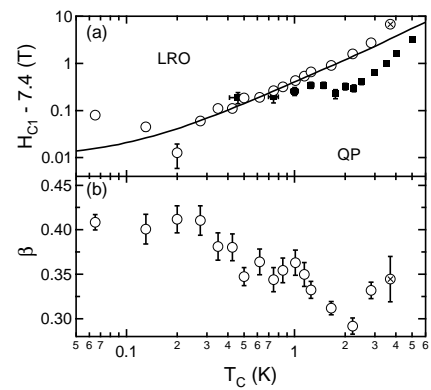} \caption{\label{fig:HcAndBeta}
(a) Phase diagram for PHCC near $H_{c1}$. Solid squares: onset of
high susceptibility state. Open circles: onset of AFM LRO from fits
described in the text. Absolute fields reported for open and closed
symbols differ by less than 0.05 T. The solid line is the calculated
mean field BEC phase boundary. (b) Critical exponent for onset of
LRO. Crossed symbols are from the $T$-dependent data, $I(T)$, in Fig.~2(b).}
\end{figure}

Figure \ref{fig:phccbraggint} shows the order parameter onset in
$H-$ and $T-$sweeps. While the onset of Bragg scattering coincides
with the onset of elevated $\chi(H)$ for $T \approx 0.4 $ K,
Bragg peaks first appear well within the high susceptibility state
for $T\approx 3$ K. A similar conclusion is reached based on the
$T-$sweep at $H=14.2$~T where the critical temperature for LRO is
$T_c(14.2\ {\rm T})=3.705(3)$ K compared to the $T \approx 8$ K
onset of the high susceptibility state. The solid line for $H < 14.2
$ T in Fig. \ref{fig:phcccontour} is the phase boundary inferred
from neutron diffraction with further details in
Fig.~\ref{fig:HcAndBeta}(a). For $T > 0.5$ K, the LRO phase resides
well within the high susceptibility state.

\begin{figure}
\centering\includegraphics[scale=0.56,trim = 0in 0.1in 0in
0in]{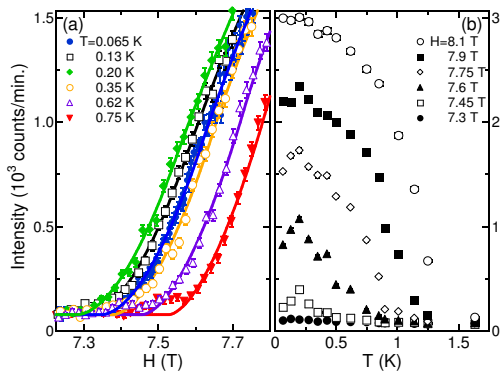}
\caption{\label{fig:phccreentrance} (Color online) Magnetic field
(a) and temperature (b) dependent scattering intensity of the
$(\frac{1}{2}0\bar{\frac{1}{2}})$ AFM Bragg peak, showing reentrant
behavior of the gapped phase near $H_{c1}$.  Solid lines are fits to
model described in text.  Temperature dependence collected from
individual magnetic field dependent measurements with a width of
$\Delta H=0.1$ T.}
\end{figure}

At low $T$, the phase boundary to N\'{e}el order approaches the
onset of the high susceptibility state, and for $T<0.4$ K there is
no intermediate phase that can be distinguished from the
data. At lower $T$, field sweeps of the
magnetic Bragg intensity, shown in Fig.~\ref{fig:phccreentrance}(a),
indicate a minimum in the phase boundary for $T\approx 0.2$ K.
Plotted versus $T$ in Fig.~\ref{fig:phccreentrance}(b), these data
show an intensity maximum for $T\approx 0.2$ K and $7.4<H<7.9$
T indicating that PHCC passes into and then back out of the LRO
phase in this field range.

Figures~\ref{fig:phccbraggint} and \ref{fig:phccreentrance} show a
rounded onset of magnetic scattering in PHCC. If critical
fluctuations are responsible for this, the energy scale must be less
than the $\approx 50~\mu\mathrm{eV}$ energy resolution.
Alternatively, field inhomogeneity can smear a singular onset. The
solids lines in Figs.~\ref{fig:phccbraggint} and
\ref{fig:phccreentrance}(a) were obtained by fitting the width of a
rectangular field distribution as well as the critical field,
$H_c(T)$, and the critical exponent, $\beta$. While a 3.6(1)\%
distribution width accounts for the data, it exceeds the $\approx
1$\% width expected over the sample volume in the HMI magnet. An
additional potential source of static broadening are impurities that
produce effective random fields\cite{fishmanaharony}.

Systematic values for $H_c(T)$ and $\beta$ were obtained by fitting the
data in Figs.~\ref{fig:phccbraggint} with the apparent field distribution
width fixed at 3.6\%. The corresponding phase boundary in
Fig.~\ref{fig:HcAndBeta}(a) affirms the existence of a wedge in $H-T$
space with neither LRO nor a spectral gap.  Taylor expansion of the phase
boundary about a generic point ($H_c$, $T_c$) on the line of transitions,
as follows $H_c(T)\approx H_c(T_c)+H_c^\prime(T_c)(T-T_c)$, clearly shows
that if $M(H,T)\propto(H-H_c(T))^\beta$ then $M(H_c,T)\propto(H_c^\prime(T_c)(T_c-T))^\beta$.
Hence the consistent values of $\beta$ extracted from $H-$ and $T-$ scans
at $T\approx 3.5$~K instill confidence in the experiment and analysis
(see Fig.~\ref{fig:HcAndBeta}(b)).

Recent experimental and theoretical work on interacting dimers
indicates that the phase transition to long range N\'{e}el order can
be described as a BEC of magnons \cite{affleck1990}. An applied
field drives the chemical potential for spin polarized magnons ($S_z=1$) to
zero causing BEC at sufficiently low $T$. In 2D, BEC can only occur
at $T=0$ so we associate the sharp increase in $\chi(H)$
indicated by solid points in Figs.~\ref{fig:phccbraggint}(a)
and ~\ref{fig:HcAndBeta}(a) with the
corresponding finite temperature quantum critical regime. In the
 immediate vicinity of the LRO phase boundary the critical phase is
denoted renormalized classical (RC) \cite{ChakravartyHalperinandNelson}
though there are no notable distinctions between the RC and QC regimes in
the present data.  The the renormalized critical regime
is characterized by a small
population of magnons that behave as individual particles. The finite
$T$ transition to N\'{e}el order may be BEC resulting from weak
inter-bi-layer coupling or a 2D Kosterlitz-Thouless (KT) transition. To
distinguish these scenarios we explore the corresponding theoretical
phase boundaries. Following Nikuni {\em et al.} \cite{Nikuni00} and
Misguich and Oshikawa \cite{Misguich04} we treat magnons as bosons
with a chemical potential $\mu=g\mu_B(H-H_c(0))$ and short range
repulsion, $v_0$:
\begin{equation}
\mathcal{H} = \sum_\mathbf{k} (\epsilon_\mathbf{k}-\mu)
a^\dagger_\mathbf{k} a_\mathbf{k} + \frac{v_0}{V}
\sum_{\mathbf{q,k,k'}} a^\dagger_\mathbf{q+k'}
a^\dagger_\mathbf{q-k'} a_\mathbf{q-k} a_\mathbf{q+k}.
\end{equation}
Mean field theory yields a condensate magnon density
$n=M/M_{sat}=g\mu_B(H-H_c(0))/2v_0$. Beyond a cusp that may be
associated with logarithmic corrections \cite{Sachdev94}, Fig. 2(a)
shows that the low $T$ $\chi(H)$ indeed displays a
plateau from which we obtain $v_0/V=1.9$~ meV, where $V$ is the unit cell volume. As expected for hard
core bosons, this number is similar to the magnon bandwidth $W=1.8$
meV.

The Hartree Fock approximation provides the effective
Hamiltonian
\begin{equation}
\mathcal{H} = \sum_\mathbf{k} (\epsilon_\mathbf{k}-\mu+v_0 n)
a^\dagger_\mathbf{k} a_\mathbf{k}. \label{eq-H}
\end{equation}
Bosons condense when the renormalized chemical potential
$\tilde{\mu} \equiv \mu - 2v_0 n_c=0$, which yields the critical
density and field
\begin{eqnarray}
n_c(T) &=& \frac{1}{V} \sum_\mathbf{k}
\frac{1}{\exp{({\epsilon_\mathbf{k}/T})}-1} \label{eq-nc}
\\H_{c}(T) &=& H_c(0) + 2v_0 n_c(T)/g \mu_B. \label{eq-Hc}
\end{eqnarray}
We assume quasi-2D magnon dispersion $\epsilon_{\bf k}=\epsilon_{\bf
k}^{2D}+2\gamma (1-\cos (k_yb))$ with $\epsilon_{\bf k}^{2D}$ from
experiments \cite{stonephccprb01}. When $T \ll \gamma$, only the
bottom of the magnon band is thermally excited and one may replace
the exact band structure with parabolic dispersion to obtain
\begin{equation}
n_c(T) = \zeta(3/2) \, \left(\frac{m^\mathrm{3D} T}{2\pi
\hbar^2}\right)^{3/2},\label{3DBEC}
\end{equation}
where $m^\mathrm{3D} = (m_a m_b m_c)^{1/3}$ is the 3D effective
mass. In the limit of a very weak inter-bi-layer tunnelling there is
a regime $\gamma \ll T \ll W$ where the in-plane dispersion can be
treated as parabolic and the critical density for a quasi-2D Bose
gas is obtained\cite{Micnas90}
\begin{equation}
n_c(T) = \frac{m^\mathrm{2D} T}{2\pi \hbar^2 b} \, \log{\frac{2
T}{\gamma}}.
\end{equation}
Here $m^\mathrm{2D} = (m_a m_c)^{1/2}$ and $b$ is the inter-bi-layer
distance.  In the intermediate regime $T = \mathcal{O}(\gamma)$ the
critical density first rises {\em faster} than $T^{3/2}$
\cite{Misguich04} before crossing over to $T \log{T}$ behavior. For
$\gamma=0.03$ meV the calculated phase boundary shown in
Fig.~\ref{fig:HcAndBeta}(a) is consistent with the data over one
decade of $T$. If bi-layers in PHCC were fully decoupled from each
other the BEC would change into a KT vortex-unbinding
transition. In 2D the crossover exponent $\phi
= 1$, so in contrast to the observed phase boundary a KT phase
boundary $H_c(T) = H_{c}(0) + C T^{\phi}$ would be linear for
$T\rightarrow 0$. This is consistent with a recent comprehensive
analysis of magnon condensation in 2D by Sachdev and Dunkel \cite{dunksach}.
Hence it appears that 3D BEC rather than vortex unbinding is the
appropriate description of the field induced transition to LRO in
PHCC.

The experimental high $T$ limit for the critical exponent
$\beta=0.34(2)$ obtained by averaging PHCC data for 0.5~K$<T<$4~K is
consistent with a 3D XY model for which $\beta=0.345$
\cite{Collins}. Upon cooling through the temperature $T\approx
0.4$~K where $H_c(T)$ merges with the 2D BEC cross over inferred
from magnetization data, the experimental values for $\beta$
increase. The apparent increase of the exponent $\beta$
is consistent with an expected crossover
from a thermally driven transition to a quantum phase transition.  Because
the upper critical dimension of the zero-temperature BEC is $d_c = 2$
\cite{Fisher89}, $\beta$ has a mean-field value of $1/2$.

A discrepancy in the description of the phase diagram for PHCC
presented so far exists for $T<0.3$ K where the observed critical
field exceeds the BEC phase boundary (Eqs.~\ref{eq-Hc} and ~\ref{eq-nc}) with a finite
$T\approx 0.2$~K minimum (Figs.~\ref{fig:HcAndBeta} and \ref{fig:phccreentrance}).
Various low energy aspects of PHCC may be
responsible for this behavior. Owing to the low symmetry of the
lattice, exchange interactions in PHCC must be anisotropic which
could lead to an Ising transition at sufficiently low $T$.
Alternatively nuclear spins, and phonons which are effectively
decoupled from magnetism at high $T$ and normally unimportant
compared to exchange interactions at low $T$ may become relevant
close to the field tuned QCP. Similar low $T$ anomalies have
been found in other electronic spin systems close to quantum
criticality such as GGG \cite{tsuiprl1999}, $\rm LiHoF_4$
\cite{bitko}, and $\rm ZnCr_2O_4$ \cite{ZincCrRef}. In $\rm
LiHoF_4$, the anomaly favors the spin ordered phase and is
associated with hyperfine coupling to the nuclear spin system. The
spin ordered phase is also favored for $\rm ZnCr_2O_4$ where the
anomaly is associated with magneto-elastic coupling.  Low
temperature spin-lattice coupling is also observed in the spin-gap
systems TlCuCl$_3$ \cite{Vyaselev04} and CuHpCl \cite{Lorenzo04}.  For PHCC, the singlet ground state may
be affected by coupling to Cu nuclear spins for $T < 0.2$ K.  This could help to
stabilize bond order over spin order and explain our failure to
discover additional phase boundaries at low $T$. Alternatively,
$H=7.4$ T and $T=0.2$ K may be a tetra-critical point separating the
bond ordered phase, the $(0.5,0,0.5)$ type spin ordered phase and a
yet to be detected magneto-elastic or nuclear+electronic spin
ordered phase.

The $H-T$ phase diagram for PHCC illustrates many important aspects
of strongly correlated systems. There is evidence for a finite $T$
crossover to a quasi-2D RC phase with 3D BEC at lower $T$ and
higher $H$. We also presented evidence for a non-monotonic phase
boundary to spin order at low $T$, which indicates that exchange
anisotropy, nuclear spin and/or lattice degrees of freedom can be
important close to quantum criticality.

We gratefully acknowledge discussions with A. Aharony, L. Balents,
O. Entin-Wohlmann, A. B. Harris,  S. Sachdev, and T. Yildrim. Work
at JHU was supported by the NSF through DMR-0074571, DMR-0306940, DMR-0348679
and by the BSF through grant No. 2000-073. ORNL is managed for the
US DOE by UT-Battelle Inc. under contract DE-AC05-00OR2272.

\end{document}